\documentclass[prd,showpacs,preprintnumbers,amsmath,amssymb]{revtex4}
\usepackage{cancel}
\usepackage{booktabs}
\usepackage{multirow}
\usepackage{slashed}
\usepackage{graphicx}
\usepackage[colorlinks,
            citecolor=blue,
            anchorcolor=red,
            menucolor=red,
            linkcolor=red,
            filecolor=red,
            runcolor=red,
            urlcolor=blue,
            frenchlinks=red]{hyperref}

\begin{document}
 \title{$X(4140)$, $X(4270)$, $X(4500)$, and $X(4700)$ and their $cs\bar{c}\bar{s}$ tetraquark partners}
\author{Jing Wu$^1$, Yan-Rui Liu$^1$}
\email{yrliu@sdu.edu.cn} \affiliation{ $^1$School of Physics and Key
Laboratory of Particle Physics and Particle Irradiation (MOE),
Shandong University, Jinan 250100, China}
\author{Kan Chen$^{2,3}$, Xiang Liu$^{2,3}$}
\email{xiangliu@lzu.edu.cn} \affiliation{
$^2$School of Physical Science and Technology, Lanzhou University, Lanzhou 730000, China\\
$^3$Research Center for Hadron and CSR Physics, Lanzhou University
and Institute of Modern Physics of CAS, Lanzhou 730000, China }
\author{Shi-Lin Zhu$^{4,5,6}$}
\email{zhusl@pku.edu.cn} \affiliation{ $^4$School of Physics and
State Key Laboratory of Nuclear Physics and Technology, Peking
University, Beijing 100871, China
\\
$^5$Collaborative Innovation Center of Quantum Matter, Beijing
100871, China
\\
$^6$Center of High Energy Physics, Peking University, Beijing
100871, China }
\begin{abstract}
In the simple color-magnetic interaction model, we investigate
possible ground $cs\bar{c}\bar{s}$ tetraquark states in the
diquark-antidiquark basis. We use several methods to estimate the
mass spectrum and discuss possible assignment for the $X$ states
observed in the $J/\psi\phi$ channel. We find that assigning the
Belle $X(4350)$ as a $0^{++}$ tetraquark is consistent with the
tetraquark interpretation for the $X(4140)$ and $X(4270)$ while the
interpretation of the $X(4500)$ and $X(4700)$ needs orbital or
radial excitation. There probably exist several tetraquarks around
4.3 GeV that decay into $J/\psi\phi$ or $\eta_c\phi$.
\end{abstract}
\pacs{14.40.Rt, 12.39.Jh}

\date{\today}

\maketitle
\section{Introduction}\label{sec1}

Recently, several exotic resonances were observed in the invariant
mass distributions of $J/\psi\phi$. In the decay process
$B^+\rightarrow J/\psi\phi K^+$, the CDF Collaboration found the
first evidence of a narrow structure X(4140) with mass
$M=4143.0\pm2.9\pm1.2$ MeV and width
$\Gamma=11.7^{+8.3}_{-5.0}\pm3.7$ MeV \cite{Aaltonen:2009tz}. Later,
the CMS Collaboration \cite{Chatrchyan:2013dma} and the D0
Collaboration \cite{Abazov:2013xda} observed structures consistent
with the $X(4140)$ in the same process. The D0 Collaboration also
observed the structure in $\bar{p}p\to J/\psi\phi+{\text anything}$
\cite{Abazov:2015sxa}. However, the Belle \cite{ChengPing:2009vu}
and $BABAR$ \cite{Lees:2014lra} experiments gave negative results for
this state in the $B$ decays. The evidence of a second $X(4274)$
resonance with mass $M=4274.4^{+8.4}_{-6.7}\pm 1.9$ MeV and width
$\Gamma=32.3^{+21.9}_{-15.3}\pm7.6$ MeV was also found in
$B^+\rightarrow J/\psi\phi K^+$ by the CDF Collaboration
\cite{Aaltonen:2011at}, which was not confirmed by the $BABAR$
Collaboration \cite{Lees:2014lra}. In the $\gamma\gamma\to
J/\psi\phi$ process, the Belle Collaboration found the evidence of a
narrow state $X(4350)$ \cite{Shen:2009vs}. The $B^+\rightarrow
J/\psi\phi K^+$ data from D\O\, also accommodate this structure
\cite{Abazov:2013xda}. In addition, the CMS Collaboration reported
the evidence of a state with mass $M=4313.8\pm5.3\pm7.3$ MeV and
$\Gamma=38^{+30}_{-15}\pm15$ MeV in the $B^+\rightarrow J/\psi\phi
K^+$ decay \cite{Chatrchyan:2013dma}. Interested readers may consult
the recent review \cite{Chen:2016qju}.

Very recently, in the $B^+\rightarrow J/\psi\phi K^+$ decay, the
LHCb Collaboration confirmed the existence of the $X(4140)$ and
$X(4274)$. Their quantum numbers are measured to be $J^{PC}=1^{++}$
\cite{Aaij:2016iza}. The mass of the $X(4140)$,
$M=4146.5\pm4.5^{+4.6}_{-2.8}$ MeV, is consistent with the world
average $M=4143.4\pm1.9$ MeV, but the width
$\Gamma=83\pm21^{+21}_{-14}$ MeV is larger than the existing value
$\Gamma=15.7\pm6.3$ MeV. In the same process, the collaboration
observed two additional higher resonances, $X(4500)$ and $X(4700)$.
Their masses and widths are $M=4506\pm11^{+12}_{-15}$ MeV,
$\Gamma=92\pm21^{+21}_{-20}$ MeV and $M=4704\pm10^{+14}_{-24}$ MeV,
$\Gamma=120\pm31^{+42}_{-33}$ MeV, respectively. Their quantum
numbers are $J^{PC}=0^{++}$.

It is difficult to understand these $X$ states in the conventional
quark-antiquark picture because their decays are expected to be
dominated by open charm channels. The proposed theoretical
explanations for the $X(4140)$ and the $X(4274)$ include molecules
\cite{Liu:2008tn,Liu:2009ei,Mahajan:2009pj,Branz:2009yt,Ding:2009vd,Liu:2010hf,Wang:2011uk,Finazzo:2011he,He:2011ed,HidalgoDuque:2012pq},
compact tetraquark states
\cite{Drenska:2009cd,Stancu:2009ka,Patel:2014vua,Lebed:2016yvr}, dynamically
generated resonances \cite{Molina:2009ct,Branz:2010rj}, and coupled
channel effects \cite{Danilkin:2009hr,vanBeveren:2009dc}. It seems
difficult to interpret them with the molecule and cusp scenarios
because of the $J^{PC}$ quantum numbers. Compared with the
tetraquark configuration, the number of the meson states is reduced
by half in the molecule configuration since the hidden-color
components are ignored. One may turn to the tetraquark picture to
understand consistently both the number and the masses of the
observed states.

From the calculation with the QCD sum rule in Ref. \cite{Chen:2016oma},
one may assign the $X(4500)$ and $X(4700)$ as the two $D$-wave
$cs\bar{c}\bar{s}$ tetraquark states of $J^P=0^+$. In another
analysis \cite{Wang:2016gxp,Wang:2016tzr}, the former state is
assigned as the first radially excited state of the
$[cs]_{1^+}[\bar{c}\bar{s}]_{1^+}$ tetraquark and the latter one as
the ground $[cs]_{1^-}[\bar{c}\bar{s}]_{1^-}$ tetraquark, but the
assignment of $X(4140)$ as a $1^{++}$ diquark-antidiquark meson is
disfavored. In Ref. \cite{Maiani:2016wlq}, Maiani $et. al$. proposed
that the $X(4140)$ and the $X(4274)$ belong to the ground state
$1S$-multiplet of diquark-antidiquark tetraquarks while the
$X(4500)$ and the $X(4700)$ are radially excited $2S$ states. Since
their tetraquark model allows only one $1^{++}$ state, the quantum
numbers of the $X(4274)$ are proposed to be $0^{++}$ or $2^{++}$. In
Ref. \cite{Zhu:2016arf}, the hidden charm tetraquarks are
investigated systematically in a diquark-antidiquark model, where
the $X(4140)$ and $X(4274)$ were explained as the $J^P=1^+$ hidden
charm tetraquarks with quark content
$\frac{1}{\sqrt6}(u\bar{u}+d\bar{d}-2s\bar{s})c\bar{c}$. The scalar
$X(4700)$ may be explained as the radial excitation of the hidden
charm tetraquarks with the same quark content while the $X(4500)$ is
its flavor singlet partner. From a potential quark model calculation
with the diquark-antidiquark picture \cite{Lu:2016cwr}, the
$X(4140)$ [$X(4700)$] can be assigned as the ground (2S excited)
tetraquark state. The $X(4500)$ can be explained as a tetraquark
composed of one $2S$ scalar diquark and one scalar antidiquark,
while the $X(4274)$ is a good candidate of the $\chi_{c1}(3P)$
charmonium. A rescattering mechanism is used in Ref.
\cite{Liu:2016onn} to understand the structure of these four $X$
states. This mechanism may explain the $X(4140)$ and $X(4700)$, but
fails to generate the $X(4274)$ and $X(4500)$, which leads to the
proposal that they are genuine resonances, e.g., $\chi_{c1}(3P)$. In
a coupled-channel quark model calculation, the authors of Ref.
\cite{Ortega:2016hde} find that the $X(4140)$ appears as a cusp
while the $X(4274)$, $X(4500)$, and $X(4700)$ appear as $3^3P_1$,
$4^3P_0$, and $5^3P_0$ charmonium states, respectively.

To understand the structures of the $X$ states decaying to
$J/\psi\phi$, we investigate systematically the mass spectrum
of the $S$-wave $cs\bar{c}\bar{s}$ system with the chromomagnetic
interaction in this work. We consider several schemes when
estimating their masses. The paper is organized as follows. In Sec.
\ref{sec2}, we present the $C$-parity eigenfunctions of different
quantum numbers and give the matrices for the chromomagnetic
interaction (CMI). In Sec. \ref{sec3}, we extract the needed
parameters and list the numerical results of the CMI matrices and
the mass spectra with different methods. Section \ref{sec4} is a short
summary.

\section{Formalism}\label{sec2}\label{sec2}

We adopt the diquark-antidiquark basis to analyze the $S$-wave
$cs\bar{c}\bar{s}$ system, where the color wave function of the
diquark belongs to the $\bar{3}_c$ or $6_c$ representation and the
antidiquark belongs to $3_c$ or $\bar{6}_c$. In the spin space, both
the diquark and the antidiquark can be a singlet or triplet state. When we say diquark in our study, it only means two quarks and the notation is convenient for us to organize the wave functions. The meaning is different from that in the diquark model \cite{Maiani:2004vq}, where the diquark is a strongly correlated quark-quark substructure with color=$\bar{3}$ and spin=0.

For the scalar and tensor states, we have
\begin{equation}
\begin{split}
 &J^{PC}=2^{++}\quad
  \phi_1\chi_1=[(cs)^1_{\bar{3}}(\bar{c}\bar{s})^1_3]^2\quad \phi_2\chi_1=[(cs)^1_{6}(\bar{c}\bar{s})^1_{\bar{6}}]^2,\\
 &J^{PC}=0^{++}\quad
  \phi_1\chi_3=[(cs)^1_{\bar{3}}(\bar{c}\bar{s})^1_3]^0\quad \phi_2\chi_3=[(cs)^1_{6}(\bar{c}\bar{s})^1_{\bar{6}}]^0\\
  &\quad\quad\quad\quad\quad\quad
  \phi_1\chi_6=[(cs)^0_{\bar{3}}(\bar{c}\bar{s})^0_3]^0\quad \phi_2\chi_6=[(cs)^0_{6}(\bar{c}\bar{s})^0_{\bar{6}}]^0,
\end{split}
\end{equation}
where the superscripts on the right side of the equation denote the
spin and the subscripts the $SU(3)_c$ representation. The notation
$\phi_1$ ($\phi_2)$ represents the color wave function with the
configuration $[\bar{3}]_c\otimes[3]_c$ ($[6]_c\otimes[\bar{6}]_c$).
The $\chi_i$ ($i=1,\cdots,6$) indicate different configurations for
the total spin wave functions coupled with the diquark and the
antidiquark. As for the axial vector tetraquarks,
  one can construct two types of eigenstates with opposite $C$ parities. Two bases naturally have negative $C$ parity,
\begin{equation}
\begin{split}
&J^{PC}=1^{+-}\quad
  \phi_1\chi_2=[(cs)^1_{\bar{3}}(\bar{c}\bar{s})^1_3]^1\quad \phi_2\chi_2=[(cs)^1_{6}(\bar{c}\bar{s})^1_{\bar{6}}]^1.
\end{split}
\end{equation}
Similar to Ref. \cite{Wu:2016vtq}, we may combine the four bases
\begin{eqnarray}
& \phi_1\chi_4=[(cs)^1_{\bar{3}}(\bar{c}\bar{s})^0_3]^1,\quad \phi_1\chi_5=[(cs)^0_{\bar{3}}(\bar{c}\bar{s})^1_3]^1,&\nonumber\\
& \phi_2\chi_4=[(cs)^1_{6}(\bar{c}\bar{s})^0_{\bar{6}}]^1,\quad
\phi_2\chi_5=[(cs)^0_{6}(\bar{c}\bar{s})^1_{\bar{6}}]^1,&
\end{eqnarray}
to get the $C$-parity eigenstates
\begin{equation}
\begin{split}
&J^{PC}=1^{++}\quad
  \phi_1\chi_p=\frac{1}{\sqrt{2}}(\phi_1\chi_4+\phi_1\chi_5)\quad
   \phi_2\chi_p=\frac{1}{\sqrt{2}}(\phi_2\chi_4+\phi_2\chi_5)\\
&J^{PC}=1^{+-}\quad
  \phi_1\chi_n=\frac{1}{\sqrt{2}}(\phi_1\chi_4-\phi_1\chi_5)\quad
   \phi_2\chi_n=\frac{1}{\sqrt{2}}(\phi_1\chi_4-\phi_1\chi_5).
\end{split}
\end{equation}
Here the subscript $p$ ($n$) means ``positive'' (negative)
$C$ parity.

In this paper, we adopt the simple chromomagnetic Hamiltonian to
estimate the mass spectrum of the $cs\bar{c}\bar{s}$ system, which reads
\begin{equation}\label{equation:CMI}
H=\sum_{i}m_i+H_{CM}=\sum_{i}m_i-\sum_{i<j}C_{ij}\widetilde{\lambda}_i\cdot\widetilde{\lambda}_j\sigma_i\cdot\sigma_j.
\end{equation}
Here, the effective mass $m_i$ for the $i$th constituent quark
incorporates not only the usual constituent quark mass but also the
effects of the kinetic energy, color confinement, and so on. The
effective coupling constants $C_{ij}$ reflect the strength for the
contact interaction. The $\sigma_i$ $(i=1,2,3)$ are the Pauli
matrices while $\widetilde{\lambda}_i=\lambda_i$ ($-\lambda^*_i$)
for quark (antiquark) with $\lambda_i$ ($i=1,\cdots,8$) being the
Gell-Mann matrices. With the constructed wave functions, it is not
difficult to get the matrices $\langle H_{CM}\rangle$ for different
states.

The above chromomagnetic Hamiltonian should work well in the
calculation of the mass splittings between the member states within
the same spin-flavor multiplet. If the mass of one state is known,
the masses of all the other member states within the same multiplet
can be predicted quite reliably. The above chromomagnetic
Hamiltonian can also be used to estimate the mass splittings of the
two systems when they have similar color-flavor configurations. For
example, one can use this Hamiltonian to calculate the mass
splitting of two tetraquark states quite reliably.

In the very beginning, we emphasize that the above
Hamiltonian is oversimplified. The kinetic energy and confinement
interaction are replaced by the constituent quark mass. With such a
crude approximation, there certainly exist large systematical
errors in the estimate of the overall hadron mass. However, the mass
splittings of the two hadron states remain reliable since most of
the inherent uncertainties cancel each other. The readers should keep
this point in mind.

For the tetraquark states with $J^{PC}=2^{++}$, we have
\begin{eqnarray}
\langle H_{CM}\rangle=\left(\begin{array}{cc}
\frac{4}{3}(4C_{cs}+C_{c\bar{c}}+2C_{c\bar{s}}+C_{s\bar{s}})&-2\sqrt{2}(C_{c\bar{c}}-2C_{c\bar{s}}+C_{s\bar{s}})\\
&-\frac{2}{3}(4C_{cs}-5C_{c\bar{c}}-10C_{c\bar{s}}-5C_{s\bar{s}})\\
\end{array}\right),
\end{eqnarray}
where the base is $(\phi_1\chi_1,\phi_2\chi_1)^T$.

For the $J^{PC}=1^{++}$ case,
\begin{eqnarray}
\langle H_{CM}\rangle=\left(\begin{array}{cc}
-\frac{4}{3}(4C_{cs}-C_{c\bar{c}}+2C_{c\bar{s}}-C_{s\bar{s}})&-2\sqrt{2}(C_{c\bar{c}}+2C_{c\bar{s}}+C_{s\bar{s}})\\
&\frac{2}{3}(4C_{cs}+5C_{c\bar{c}}-10C_{c\bar{s}}+5C_{s\bar{s}})\\
\end{array}\right),
\end{eqnarray}
where the base is $(\phi_1\chi_p,\phi_2\chi_p)^T$.

In the case $J^{PC}=1^{+-}$, one gets
\begin{eqnarray}
\langle H_{CM}\rangle=\left(\begin{array}{cccc}
\frac{4}{3}\left(\begin{array}{c}4C_{cs}-C_{c\bar{c}}\\-2C_{c\bar{s}}-C_{s\bar{s}}\end{array}\right)&
2\sqrt{2}(C_{c\bar{c}}-2C_{c\bar{s}}+C_{s\bar{s}})&
\frac{8}{3}(C_{c\bar{c}}-C_{s\bar{s}})&-4\sqrt{2}(C_{c\bar{c}}-C_{s\bar{s}})\\
&-\frac{2}{3}\left(\begin{array}{c}4C_{cs}+5C_{c\bar{c}}\\+10C_{c\bar{s}}+5C_{s\bar{s}}\end{array}\right)&
-4\sqrt{2}(C_{c\bar{c}}-C_{s\bar{s}})&\frac{20}{3}(C_{c\bar{c}}-C_{s\bar{s}})\\
&&-\frac{4}{3}\left(\begin{array}{c}4C_{cs}+C_{c\bar{c}}\\-2C_{c\bar{s}}+C_{s\bar{s}}\end{array}\right)&
2\sqrt{2}(C_{c\bar{c}}+2C_{c\bar{s}}+C_{s\bar{s}})\\
&&&\frac{2}{3}\left(\begin{array}{c}4C_{cs}-5C_{c\bar{c}}\\+10C_{c\bar{s}}-5C_{s\bar{s}}\end{array}\right)
\end{array}\right),
\end{eqnarray}
where the base is
$(\phi_1\chi_2,\phi_2\chi_2,\phi_1\chi_n,\phi_2\chi_n)^T$.

As for the $J^{PC}=0^{++}$ case, we have
\begin{eqnarray}
\langle H_{CM}\rangle=\left(\begin{array}{cccc}
\frac{8}{3}\left(\begin{array}{c}2C_{cs}-C_{c\bar{c}}\\-2C_{c\bar{s}}-C_{s\bar{s}}\end{array}\right)&
4\sqrt{2}\left(\begin{array}{c}C_{c\bar{c}}-2C_{c\bar{s}}\\+C_{s\bar{s}}\end{array}\right)&
-\frac{4}{\sqrt{3}}(C_{c\bar{c}}-2C_{c\bar{s}}+C_{s\bar{s}})&2\sqrt{6}(C_{c\bar{c}}+2C_{c\bar{s}}+C_{s\bar{s}})\\
&-\frac{4}{3}\left(\begin{array}{c}2C_{cs}+5C_{c\bar{c}}\\+10C_{c\bar{s}}+5C_{s\bar{s}}\end{array}\right)&
2\sqrt{6}(C_{c\bar{c}}+2C_{c\bar{s}}+C_{s\bar{s}})&\frac{-10}{\sqrt{3}}(C_{c\bar{c}}-2C_{c\bar{s}}+C_{s\bar{s}})\\
&&-16C_{cs}&0\\
&&&8C_{cs}
\end{array}\right),
\end{eqnarray}
where the base is
$(\phi_1\chi_3,\phi_2\chi_3,\phi_1\chi_6,\phi_2\chi_6)^T$.

When deriving these matrices, we have adopted the approach used in Refs. \cite{Hogaasen:2004pm,Buccella:2006fn}. The spin and color matrix elements are calculated separately with $H_S=\sum_{i<j}C_{ij}\sigma_i\cdot\sigma_j$ and $H_C=-\sum_{i<j}C_{ij}\widetilde{\lambda}_i\cdot\widetilde{\lambda}_j$, respectively. Then one performs a type of ``tensor product'' of $\langle H_S\rangle$ and $\langle H_C\rangle$ to get the final $\langle H_{CM}\rangle$. For example, if one obtains $\langle \chi_x|H_S|\chi_y\rangle=a_S C_{12}+ b_S C_{13}+\cdots$ and $\langle \phi_\alpha|H_S|\phi_\beta\rangle=a_C C_{12}+ b_C C_{13}+\cdots$, one gets the color-magnetic matrix element of $\langle \phi_\alpha\chi_x |H_{CM}|\phi_\beta\chi_y\rangle$ by multiplying coefficients of corresponding coupling constants: $\langle \phi_\alpha\chi_x |H_{CM}|\phi_\beta\chi_y\rangle=(a_S*a_C) C_{12}+ (b_S*b_C) C_{13}+\cdots$. The matrix elements in spin space are easy to calculate \cite{Maiani:2004vq}. We here give those in color space
\begin{eqnarray}
\left(\begin{array}{cc}
\langle\phi_1| H_C |\phi_1\rangle & \langle\phi_1| H_C |\phi_2\rangle\\
&\langle\phi_2| H_C|\phi_2\rangle
\end{array}\right)
&=&
\left(\begin{array}{cc}
\frac43(4C_{cs}+2C_{c\bar{s}}+C_{c\bar{c}}+C_{s\bar{s}})&2\sqrt{2}(2C_{c\bar{s}}-C_{c\bar{c}}-C_{s\bar{s}})\\
&\frac23(-4C_{cs}+10C_{c\bar{s}}+5C_{c\bar{c}}+5C_{s\bar{s}})
\end{array}\right).
\end{eqnarray}
With this matrix, it is easy to check the consistency between our formulas and those in the diquark model \cite{Maiani:2004vq}.

\section{Model parameters and numerical results}\label{sec3}

To estimate the masses of these tetraquark states, we need to
extract six parameters: the effective masses $m_c$ and $m_s$,
and effective coupling constants $C_{cs}$, $C_{c\bar{s}}$,
$C_{c\bar{c}}$, and $C_{s\bar{s}}$. They may be extracted from the
known baryons and mesons, where we have assumed that they do not
change much from system to system.

The coupling constants rely only on the mass splittings of hadrons.
We summarize the adopted hadrons and the obtained coupling constants
in Table \ref{parameter}. When extracting $C_{s\bar{s}}$, one has to
use the mass of the ground pseudoscalar meson. Since it is affected
significantly by chiral symmetry, we use $C_{s\bar{s}}=C_{ss}=6.4$
MeV for the calculation. This value is determined through
$2m_{\Omega^-}+m_\Delta-(2m_{\Xi^{*0}}+m_\Xi)=8C_{ss}+8C_{qq}$. With
the above parameters, one gets the numerical values for the $\langle
H_{CM}\rangle$ matrices and their eigenvalues and eigenvectors. The
results are collected in Table \ref{table:cscs}.

\begin{table}[h!]
\caption{Color-magnetic interaction ($\langle H_{CM}\rangle$) for
various hadrons and the obtained effective coupling constants in
units of MeV through the mass differences between the hadrons in the
two columns.}\label{parameter} \centering
\begin{tabular}{cccccc}
\hline\hline
Hadron&CMI&Hadron&CMI&Value\\
\hline
$N$&$-8C_{qq}$&$\Delta$&$8C_{qq}$&$C_{qq}=18.4$\\
$\Sigma$&$\frac{8}{3}C_{qq}-\frac{32}{3}C_{qs}$&$\Sigma^*$&$\frac{8}{3}C_{qq}+\frac{16}{3}C_{qs}$&$C_{qs}=12.4$\\
$D$&$-16C_{c\bar{q}}$&$D^{*}$&$\frac{16}{3}C_{c\bar{q}}$&$C_{c\bar{q}}=6.8$\\
$D_s$&$-16C_{c\bar{s}}$&$D_{s}^{*}$&$\frac{16}{3}C_{c\bar{s}}$&$C_{c\bar{s}}=6.8$\\
$\eta_{c}$&$-16C_{c\bar{c}}$&$J/\psi$&$\frac{16}{3}C_{c\bar{c}}$&$C_{c\bar{c}}=5.3$\\
$\Sigma_{c}$&$\frac{8}{3}C_{qq}-\frac{32}{3}C_{qc}$&$\Sigma_{c}^*$&$\frac{8}{3}C_{qq}+\frac{16}{3}C_{qc}$&$C_{cq}=4.0$\\
$\Xi'_{c}$&$\frac{8}{3}C_{qs}-\frac{16}{3}C_{qc}-\frac{16}{3}C_{sc}$&$\Xi_{c}^*$&$\frac{8}{3}C_{qs}+\frac{8}{3}C_{qc}+\frac{8}{3}C_{sc}$&$C_{sc}=4.6$\\
\hline
\end{tabular}
\end{table}

\begin{table}[h!]
\caption{Color-magnetic interactions for the $cs\bar{c}\bar{s}$
system in units of MeV.}\label{table:cscs} \centering
\begin{tabular}{c|ccc}
\hline
$J^{PC}$ & $\langle H_{CM}\rangle$& Eigenvalue&Eigenvector \\
\hline $2^{++}$
&$\left(\begin{array}{cc}58.3&5.4\\&72.1\end{array}\right)$
&$\left(\begin{array}{c}73.9\\56.4\end{array}\right)$
&$\left(\begin{array}{c}0.32,0.95\\-0.95,0.32\end{array}\right)$
\\
$1^{++}$
&$\left(\begin{array}{cc}-27.1&-71.6\\&5.9\end{array}\right)$
&$\left(\begin{array}{c}-84.0\\62.9\end{array}\right)$
&$\left(\begin{array}{c}-0.78,-0.62\\0.62,-0.78\end{array}\right)$
\\
$1^{+-}$
&$\left(\begin{array}{cccc}-9.2&-5.4&-2.9&6.2\\&-96.6&6.2&-7.3\\&&-22.0&71.6\\&&&18.6\end{array}\right)$
&$\left(\begin{array}{c}-100.4\\-73.3\\72.8\\-8.3\end{array}\right)$
&$\left(\begin{array}{c}-0.03,-0.93,0.28,-0.22\\-0.12,-0.35,-0.74,0.56\\
-0.04,0.01,-0.60,-0.80\\0.99,-0.07,-0.10,0.03\end{array}\right)$
\\
$0^{++}$
&$\left(\begin{array}{cccc}-42.9&-10.7&4.4&123.9\\&-180.9&123.9&11.0\\&&-73.6&0\\&&&36.8\end{array}\right)$
  &$\left(\begin{array}{c}-264.0\\-131.8\\127.2\\7.9\end{array}\right)$
  &$\left(\begin{array}{c}-0.09,-0.83,0.54,0.07\\-0.80,0.07,-0.09,0.59\\
  -0.59,-0.01,-0.02,-0.81\\0.05,-0.55,-0.83, 0.00\end{array}\right)$\\\hline
\end{tabular}
\end{table}

We can evaluate the mass spectrum of the $cs\bar{c}\bar{s}$ system
if we know the effective masses $m_i$. Since they incorporate the
quark kinetic energy and confinement effects, in principle, their
values are different for various systems and cannot be determined
uniformly. However, as a rough estimation, we would extract the
effective quark masses from the known baryons. For example, the
color-magnetic interaction for the nucleon is $\langle
H_{CM}\rangle$=$-8C_{qq}$ ($q=u,d$). From the mass formula
$M=\sum_{i}m_i+\langle H_{CM}\rangle$, one gets $m_q=361.8$ MeV.
Similarly, we get $m_s=540.4$ MeV from $M_\Omega=3m_s+8C_{ss}$. With
the values of the coefficients $C_{qq}$ and $C_{cq}$ and the formula
$m_c=(3M_{\Sigma^*_c}-2M_{\Delta}-16C_{qc}+8C_{qq})/3$, one obtains
$m_c=1724.8$ MeV.

Before estimating the masses of the $cs\bar{c}\bar{s}$ system, we
take a look at the conventional hadrons with the determined
parameters. The calculated masses are listed in Table \ref{text}.
From the values, it is obvious that the obtained hadron masses are
larger than the experimental data. The discrepancy can even reach
377 MeV for the mesons. Therefore, the resultant estimations with
these effective masses should be taken as a theoretical upper limit.
\begin{table}[h!]
\caption{Comparison for hadron masses between experimental data and
theoretical estimation. All the values are in units of
MeV.}\label{text} \centering
\begin{tabular}{c|ccc|c|ccc}
\hline
Hadron &Theory&Experiment&Deviation&Hadron &Theory &Experiment &Deviation \\
\hline
$D$&1975.9&1864.8&111.1&$D^*$&2121.0&2007.0&114.0\\
\hline
$D_s$&2154.5&1968.3&186.2&$D_s^*$&2299.5&2112.1&187.4\\
\hline
$\eta_c$&3361.0&2983.6&377.4&$J/\psi$&3474.1&3096.9&377.2\\
\hline
$\Sigma_c$&2452.9&2454.0&1.1&$\Sigma_c^*$&2516.9&2518.4&-1.5\\
\hline
$\Omega_c$&2796.2&2695.2&101.0&$\Omega_c^*$&2845.3&2765.9&79.4\\
\hline
$\Xi_c$&2525.9&2471.0&54.9&$\Xi_c^{'}$&2612.3&2577.9&34.4\\
\hline
$\Xi_c^{*}$&2680.6&2645.9&34.7&\\
\hline
\end{tabular}
\end{table}

We here use two methods to discuss the tetraquark masses: (1)
substitute all the obtained parameters into the formula
$M=\sum_{i}m_i+\langle H_{CM}\rangle$; and (2) estimate the results
with some reference parameter, i.e. $M-M_{ref}=\langle
H_{CM}\rangle-\langle H_{CM}\rangle_{ref}$. It is not necessary to
use the effective quark masses with the latter method, where the
quark mass effects are partly eliminated. In the present study, one
estimates the tetraquark masses with three reference parameters, the
threshold of $D_s^+D_s^{*-}$, the threshold of $J/\psi\phi$, and the
mass of the $Y(4140)$.

\begin{table}[h!]
\caption{Mass spectrum of the $cs\bar{c}\bar{s}$ system in the
effective quark mass method in units of MeV.}\label{table:mass
spectrum1} \centering
\begin{tabular}{c|cccc}
\hline
$J^{PC}$ &\multicolumn{4}{c}{Tetraquark mass} \\
\hline
$2^{++}$&4600.5&4583.0\\
$1^{++}$&4589.5&4442.6\\
$1^{+-}$&4599.4&4518.3&4453.3&4426.3\\
$0^{++}$&4653.8&4534.5&4394.8&4262.6\\\hline
\end{tabular}
\end{table}
In the effective quark mass method, the mass spectrum for the
$cs\bar{c}\bar{s}$ system is given in Table \ref{table:mass
spectrum1}. The highest and the lowest tetraquarks are both scalars.
Although the two highest masses 4654 and 4533 MeV in the
$J^{PC}=0^{++}$ case are not far from the observed $X(4700)$ and
$X(4500)$, it is difficult to assign the observed scalars as ground
tetraquarks since our results are overestimated numbers. The two
$J^{PC}=1^{++}$ tetraquarks are also 300 MeV higher than the
$X(4140)$ and $X(4274)$. However, the mass splitting between the two
$1^{++}$ tetraquarks is consistent with experiments and Stancu's
result \cite{Stancu:2009ka}. If the overestimation is 300 MeV, it is
possible to interpret the $X(4140)$ and $X(4274)$ as
$cs\bar{c}\bar{s}$ tetraquark states. As a byproduct, although the
matrices $\langle H_{CM}\rangle$ in the diquark-antidiquark basis
are different from those in Ref. \cite{Stancu:2009ka}, the
eigenvalues are the same [after correcting typos in Eq. (15) in Ref.
\cite{Stancu:2009ka}] if we use the same effective coupling
constants. So one does not need to distinguish the two pictures for
the compact $cs\bar{c}\bar{s}$ system once the diagonalization is
performed.

In the second method, we first use the threshold of $D_sD_s^*$ as a
reference parameter. The color-magnetic interaction for the
reference system reads $\langle
H_{CM}\rangle_{D_sD_s^*}=-\frac{16}{3}(3C_{c\bar{s}}-C_{c\bar{s}})=-72.5$
MeV and the mass of a tetraquark is given by the formula
$M_{tetra}=m_{D_s}+m_{D_s^*}-\langle
H_{CM}\rangle_{D_sD_s^*}+\langle H_{CM}\rangle_{tetra}$. In the
diquark-antidiquark model, one may assume a $[cs]$ substructure with
the fixed color representation $\bar{3}_c$ (or $6_c$) or consider a
general picture that the color representation can also be $6_c$ (or
$\bar{3}_c$). The resulting spectra are different. We show both
results in Table \ref{table:threshold2}. If one uses $D_sD_s$ or
$D_s^*D_s^*$ as the reference system, one gets the same results.
From the table (also Table \ref{table:cscs}), we know that the
off-diagonal matrix elements influence significantly the eigenvalues
except for the $J^{PC}=2^{++}$ case. It is obvious that the mass
splitting (33 MeV) between the two $1^{++}$ tetraquarks without
color mixing is much smaller than experiments. Therefore, one cannot
understand the two $1^{++}$ states in the ground diquark-antidiquark
picture if the color mixing is not included. If the observed mesons
are really tetraquark states, the obtained masses are around 70 MeV
lower than experimental measurements. Recall that the $H$-dibaryon
was found to be stable when one uses only the color-spin interaction
\cite{Jaffe:1976yi} while no evidence for its existence is observed.
This situation indicates that the present method needs improvement.
A possible contribution to fix this discrepancy is the additional
kinetic energy in forming a compact quark structure
\cite{Park:2015nha}.

\begin{table}[h!]
\caption{Mass spectrum of the $cs\bar{c}\bar{s}$ system in units of
MeV by using the threshold of $D_s\bar{D}^*_s$ as a reference
parameter. Part$_1$ (Part$_2$) is the case in which the mixing between
$\bar{3}_c$ and $6_c$ for the diquark $[cs]$ is (not)
considered.}\label{table:threshold2} \centering
\begin{tabular}{c|cccc|cccc}
\hline $J^{PC}$
&\multicolumn{4}{c}{Part$_{1}$}&\multicolumn{4}{c}{Part$_{2}$}
\\\hline
$2^{++}$&4226.9&4209.4&&&4225.0&4211.2\\
$1^{++}$&4215.8&4068.9&&&4158.9&4125.9\\
$1^{+-}$&4225.8&4144.6&4079.6&4052.6&4171.5&4143.7&4130.9&4056.3\\
$0^{++}$&4280.2&4160.8&4021.1&3889.0&4189.7&4110.0&4079.3&3972.0\\\hline
\end{tabular}
\end{table}

Now we evaluate the masses of the possible $cs\bar{c}\bar{s}$
tetraquak states with the $J/\psi\phi$ threshold. Since $\langle
H_{CM}\rangle_{J/\psi\phi}=\frac{16}{3}(C_{c\bar{c}}+C_{s\bar{s}})=62.4$
MeV, the obtained tetraquark masses are about 99 MeV below the
values of part$_1$ in Table \ref{table:threshold2}. This number is
from the change of the reference parameter that results in
$\delta=(m_{D_s}+m_{D_s^*}-\langle
H_{CM}\rangle_{D_sD_s^*})-(m_{J/\psi}+m_{\phi}-\langle
H_{CM}\rangle_{J/\psi\phi})\approx99$ MeV. In principle, the
tetraquark masses should not change with the choice of the reference
parameter. However, the structures of a quarknonium and a
heavy-light meson are different, and thus the kinetic energies and
the confinement strengths are different. One cannot consider such
differences in the present method. From the comparison in Table
\ref{text}, it is clear that more quark attraction for a charmonium
is needed to reproduce the experimental data than that for a
heavy-light meson. The resulting masses with the $J/\psi\phi$
threshold should be underestimated and can be treated as a
theoretical lower limit. Once the effects such as kinetic energy and
confinement are incorporated appropriately, the mass inconsistency
will be fixed.

In the diquark-antidiquark picture, the charm quark and anticharm
quark can be a color-singlet state or a color-octet state. We now
move on to the spectrum of the
$(c\bar{c})_{8_c}(s\bar{s})_{8_c}$-type tetraquarks. Their masses
can also be estimated from the $J/\psi\phi$ threshold. There are six
such state vectors: $[(c\bar{c})_8^1(s\bar{s})_8^1]^2$ with
$J^{PC}=2^{++}$, $[(c\bar{c})_8^1(s\bar{s})_8^1]^1$ with
$J^{PC}=1^{++}$, $[(c\bar{c})_8^1(s\bar{s})_8^0]^1$ and
$[(c\bar{c})_8^0(s\bar{s})_8^1]^1$ with $J^{PC}=1^{+-}$, and
$[(c\bar{c})_8^1(s\bar{s})_8^1]^0$ and
$[(c\bar{c})_8^0(s\bar{s})_8^0]^1$ with $J^{PC}=0^{++}$. Relevant
color-magnetic matrix elements are:
\begin{eqnarray}
\langle H_{CM}\rangle_{2^{++}}&=&-\frac23(C_{c\bar{c}}-4C_{cs}-14C_{c\bar{s}}+C_{s\bar{s}})=67.9\text{ MeV},\nonumber\\
\langle H_{CM}\rangle_{1^{++}}&=&-\frac23(C_{c\bar{c}}+4C_{cs}+14C_{c\bar{s}}+C_{s\bar{s}})=-83.5 \text{ MeV},\nonumber\\
\langle H_{CM}\rangle_{1^{+-}}&=&
\left(\begin{array}{cc}-\frac23(C_{c\bar{c}}-3C_{s\bar{s}})&\frac43(2C_{cs}-7C_{c\bar{s}})\\
&\frac23(3C_{c\bar{c}}-C_{s\bar{s}})\end{array}\right)=
\left(\begin{array}{cc}9.3&-51.2\\
&6.3\end{array}\right) \text{ MeV },
\nonumber\\
\langle H_{CM}\rangle_{0^{++}}&=&
\left(\begin{array}{cc}-\frac23(8C_{cs}+28C_{c\bar{s}}+C_{c\bar{c}}+C_{s\bar{s}})&-\frac{4}{\sqrt3}(2C_{cs}-7C_{c\bar{s}})\\
&2(C_{c\bar{c}}+C_{s\bar{s}})\end{array}\right)=
\left(\begin{array}{cc}-159.3&88.7\\
&23.4\end{array}\right) \text{ MeV}.
\end{eqnarray}
The eigenvalues for the $1^{+-}$ case are $59.0$ and $-43.4$
MeV. Those for the $0^{++}$ case are $-195.2$ and $59.4$ MeV. We
show the estimated masses for the $(cs)(\bar{c}\bar{s})$ tetraquarks
and $(c\bar{c})_{8_c}(s\bar{s})_{8_c}$ tetraquarks in Table
\ref{table:threshold}. They have comparable masses. If the low mass
states in the $\eta_c\phi$, $\eta_c\pi\pi\pi$, or $J/\psi\pi\pi\pi$
channel ($J/\psi\phi$ channel is closed) could be observed, the
$(c\bar{c})_{8_c}(s\bar{s})_{8_c}$ configuration is probably a more
appropriate structure.

\begin{table}[h!]
\caption{Mass spectrum of the $cs\bar{c}\bar{s}$ system in units of
MeV by using the threshold of $J/\psi\phi$ as a reference
parameter.}\label{table:threshold} \centering
\begin{tabular}{c|cccc|cccc}
\hline $J^{PC}$
&\multicolumn{4}{c}{$(cs)(\bar{c}\bar{s})$}&\multicolumn{4}{c}{$(c\bar{c})_{8_c}(s\bar{s})_{8_c}$}
\\\hline
$2^{++}$&4127.9&4110.4&&&4121.9\\
$1^{++}$&4116.9&3970.0&&&3970.4\\
$1^{+-}$&4126.8&4045.6&3980.6&3953.6&4113.0&4010.6\\
$0^{++}$&4181.2&4061.9&3922.1&3790.0&4113.4&3858.7\\\hline
\end{tabular}
\end{table}

Finally, we check the consistency for the observed mesons in the
tetraquark picture by assigning the $X(4140)$ as the lowest $1^{++}$
state. Since all these structures are compact $cs\bar{c}\bar{s}$
states, the simple chromomagnetic Hamiltonian should work well in
the calculation of their mass splittings as we emphasized in Sec.
\ref{sec2}. The resulting masses for its partners should be
relatively accurate. This approach is the most reliable one in the
calculation of the overall mass of the other $cs\bar{c}\bar{s}$
states once the $X(4140)$ is identified as the lowest $1^{++}$
state.

\begin{figure}[htpb]
\includegraphics[width=0.6\textwidth]{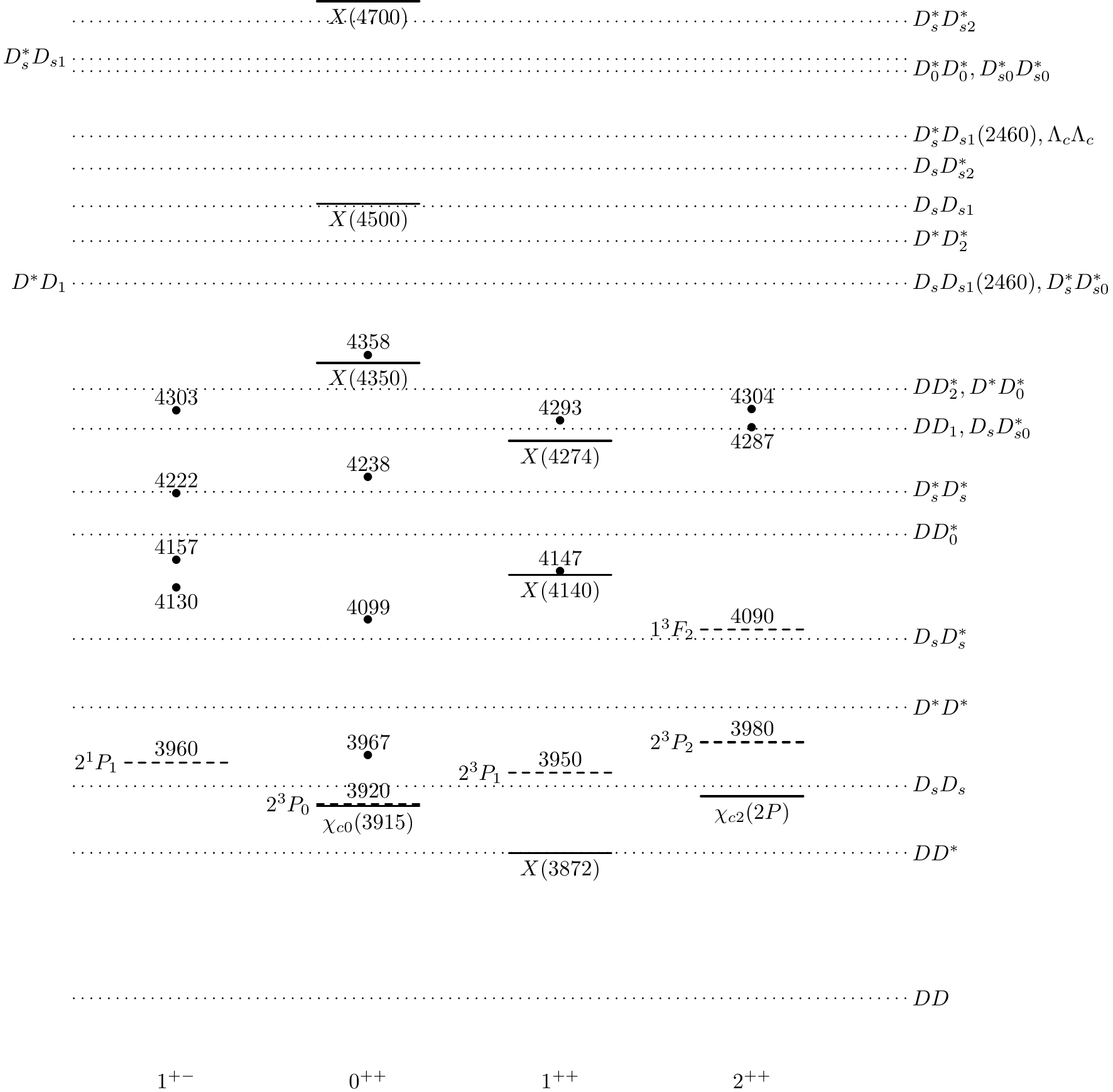}
\caption{Predicted mesons in the Godfrey-Isgur model
\cite{Godfrey:1985xj} (dashed lines), observed mesons with relevant
$J^{PC}$ (solid lines), partner states of the LHCb $X(4140)$ in the
tetraquark picture (black dots), and various thresholds (dotted
lines). The masses are given in units of MeV.}\label{fig-where}
\end{figure}

We list the obtained values in Table \ref{table:mass spectrum5}. For
comparison, we present these tetraquarks, the observed mesons with
the same quantum numbers, and the quark model predictions with
relevant $J^{PC}$ in Fig. \ref{fig-where}. Various meson-antimeson
thresholds are also shown. From the figure, the two $1^{++}$ states,
$X(4140)$ and $X(4274)$, are consistent with a tetraquark
interpretation. It is very interesting to note that the $X(4350)$
state observed by the Belle collaboration \cite{Shen:2009vs} is
consistent with the highest scalar tetraquark with the mass 4358
MeV. If the quantum numbers of this state can be confirmed to be
$J^{PC}=0^{++}$, it is very likely that more states in the
$J/\psi\phi$ or $\eta_c\phi$ invariant mass distribution could be
observed. At least four states exist around 4.3 GeV.

\begin{table}[h!]
\caption{Mass spectrum of the $cs\bar{c}\bar{s}$ system in units of
MeV by assigning the $X(4140)$ as the lowest $1^{++}$
state.}\label{table:mass spectrum5} \centering
\begin{tabular}{c|cccc}
\hline $J^{PC}$ &\multicolumn{4}{c}{Tetraquark mass} \\\hline
$2^{++}$&4304.4&4286.9\\
$1^{++}$&4293.4&4146.5\\
$1^{+-}$&4303.3&4222.2&4157.2&4130.2\\
$0^{++}$&4357.7&4238.4&4098.7&3966.5\\
\hline
\end{tabular}
\end{table}

\section{Discussions and summary}\label{sec4}

In the simple color-magnetic interaction model, we have analyzed the
spectrum of the possible ground $cs\bar{c}\bar{s}$ tetraquark
system. We use a diquark-antidiquark basis and find that the
obtained mass splittings are the same as those in the
$(c\bar{c})-(s\bar{s})$ basis \cite{Stancu:2009ka}, which indicates
that it is not necessary for us to care about the substructure for
the compact $cs\bar{c}\bar{s}$ system. However, this conclusion is
not applicable to the other compact multiquark systems, which rely
on the coupling strengths between (anti)quarks.

Because the effective quark masses contain contributions from
kinetic energy, confinement, etc., it is almost impossible to find a
universal set of values for differen hadron systems. We first tried
to estimate tetraquark masses with the effective quark masses
derived from the conventional hadrons. One gets the overestimated
results which can be treated as the upper limits of the tetraquark
masses.

In order to partly cancel the uncertainty from the quark masses, we
use the threshold of $D_sD_s^*$ (or $D_sD_s$) as a reference
parameter. The results are about 70 MeV lower than the experimental
masses. If one uses the threshold of $J/\psi\phi$, much lower masses
are obtained that can be treated as the lower limits of the
tetraquark masses. Probably the inclusion of corrections from
kinetic energy and confinement may fix the discrepancies.

If the $X(4140)$ is identified as the lowest $J^{PC}=1^{++}$
$cs\bar{c}\bar{s}$ tetraquark, we get a consistent assignment that
the $X(4274)$ could be another $1^{++}$ tetraquark and the $X(4350)$
could be the highest $0^{++}$ tetraquark.

Although the overestimation for the masses is around 300 MeV in the
first method, underestimation is around 70 MeV in the second scheme,
and underestimation is around 160 MeV in the third scheme, the mass
splittings between these $cs\bar{c}\bar{s}$ tetraquark states in
different approaches are consistent with each other.

In determining the parameters, we have used the hypothesis that the effective coupling constants in the conventional hadrons can be applied to multiquark states. One should note that whether this direct extension is appropriate needs further investigations. The reason is that the couplings are proportional to the overlap function of the two constituents, $|\psi(0)|^2$, and no principle says that the functions are the same for all types of hadrons. To investigate the spectrum of possible tetraquarks without this hypothesis, a ``type-II'' diquark was proposed in Ref. \cite{Maiani:2014aja}, where the spin-spin interaction inside the diquark is much stronger than other possible pairing and diquarks are more compact ingredients.

\begin{figure}[htpb]
\includegraphics[width=0.6\textwidth]{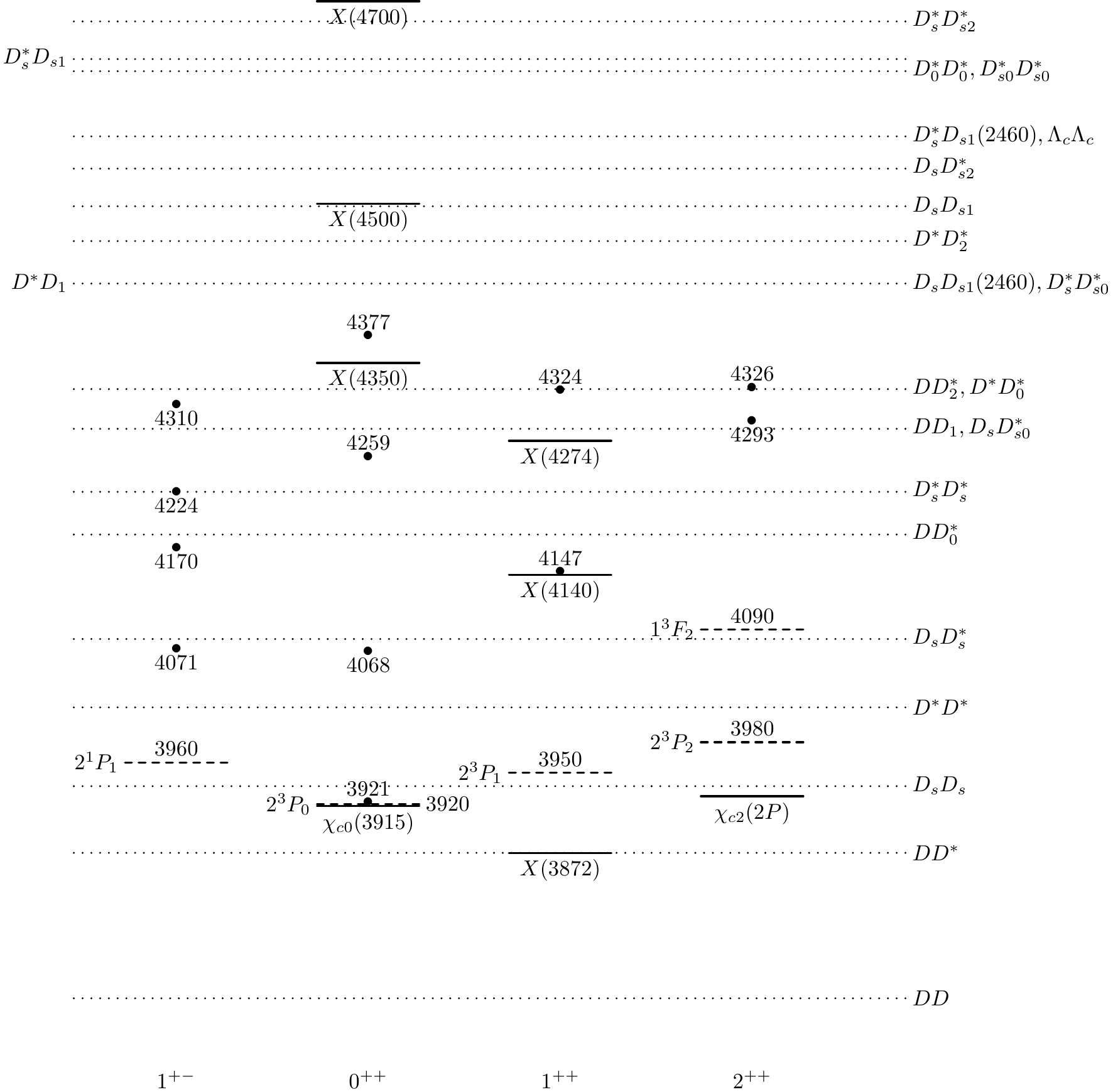}
\caption{Same as Fig. \ref{fig-where}. The masses of the partner states of the LHCb $X(4140)$ in the
tetraquark picture are estimated with $C_{cs}=4.2$ MeV, $C_{s\bar{s}}=11.3$ MeV, $C_{c\bar{s}}=6.8$ MeV, and $C_{c\bar{c}}=5.6$ MeV.}\label{fig-where2}
\end{figure}

Even with this hypothesis, the different methods to estimate the $C_{s\bar{s}}$ may lead to the slightly different tetraquark masses. With the formulaes in Sec. \ref{sec2}, one obtains the correspondence between our parameters and those used in Ref. \cite{Maiani:2004vq} is: $(\kappa_{cs})_{\bar3}\leftrightarrow\frac{16}{3}C_{cs}$ (22 MeV $\leftrightarrow$ 24.5 MeV), $\kappa_{s\bar{s}}\leftrightarrow\frac83C_{s\bar{s}}$ (30 MeV$\leftrightarrow$ 17.1 MeV),
$\kappa_{c\bar{s}}\leftrightarrow\frac83 C_{c\bar{s}}$ (18 MeV $\leftrightarrow$ 18.1 MeV), and $\kappa_{c\bar{c}}\leftrightarrow\frac83C_{c\bar{c}}$  (15 MeV $\leftrightarrow$ 14.1 MeV). If we use the parameters consistent with that work, one gets the tetraquark masses shown in Fig. \ref{fig-where2}. The choice of the coupling constants leads to the uncertainty around tens of MeV. Considering the uncertainties in both theoretical estimation and experimental measurement, the discussions in this paper are not affected largely.

To summarize, the X(4500) and X(4700) are not good candidates of the
$S$-wave $cs\bar{c}\bar{s}$ tetraquark states. A $D$-wave
excitation, two $P$-wave excitations, or a radial excitation is
needed to understand the structure of these two higher scalars. On
the other hand, the $X(4140)$, $X(4274)$, and $X(4350)$ are
consistent with the tetraquark assignment. There exist at least
three states around 4.3 GeV, which may be observed in the
$J/\psi\phi$ or $\eta_c\phi$ channel. Below the $J/\psi\phi$
threshold (4120 MeV), there may exist two scalar $cs\bar{c}\bar{s}$
tetraquark states as shown in Table \ref{table:mass spectrum5}.
These states should be narrow and can be searched for in the
$J/\psi\pi\pi\pi$ or radiative decay channels.

\section*{ACKNOWLEDGMENTS}
Y.~R.~L. thanks Professor. Su-Houng Lee for helpful discussions at the YITP
workshop YITP-W-16-01 ``MIN16 - Meson in Nucleus 2016.'' This
project is supported by the National Natural Science Foundation of China
under Grants No. 11175073. No. 11275115, No. 11222547, No. 11175073,
No. 11575008, and No. 11261130311, and the Fundamental Research Funds for
the Central Universities and 973 program. X.~L. is also supported by
the National Program for Support of Youth Top-Notch Professionals.



\end{document}